\documentstyle{aipproc}

\def\itl{\'{\i}}
\def\etal{{\it et al.~}}

\def\beq{\begin{equation}}
\def\eeq{\end{equation}}

\def\gam{{$\gamma$}}

\def\msun{{\rm ~M}_\odot}
\def\AM{{ \hat{\rm A} }}

\def\cB{{\cal B}}

\def\cM{{\cal M}}
\def\bh{{\rm bh}}
\def\cH{{\cal H}}

\def\cS{{\cal S}}

\def\Valf{{\it v}^{\rm Alf}}
\def\Alf{{\rm Alfv\'en}~}
\def\Oor{{\rm Oort}}
\def\SS{{\rm SS}}

\def\Pgas{p_{\rm gas}}

\def \ltaprx {\lower .1ex\hbox{\rlap{\raise .6ex\hbox{\hskip .3ex
	{\ifmmode{\scriptscriptstyle <}\else 
		{$\scriptscriptstyle <$}\fi}}}
	\kern -.4ex{\ifmmode{\scriptscriptstyle \sim}\else 
		{$\scriptscriptstyle\sim$}\fi}}}
\def\gtaprx {\lower .1ex\hbox{\rlap{\raise .6ex\hbox{\hskip .3ex
	{\ifmmode{\scriptscriptstyle >}\else 
		{$\scriptscriptstyle >$}\fi}}}
	\kern -.4ex{\ifmmode{\scriptscriptstyle \sim}\else 
		{$\scriptscriptstyle\sim$}\fi}}}

\def\sec{{\rm ~sec}}

\begin{document}
\thispagestyle{empty}

\title{An Instability-driven Dynamo  \\ for Gamma-ray Burst Black Holes}

\author {Rafael \'Angel 
Araya-G\'ochez\footnote{e-mail: araya@twinkie.gsfc.nasa.gov}}	
\address{Laboratorio de Investigaciones Astrof{\itl}sicas,
	 Escuela de F{\itl}sica 	\\  
  	 Universidad de Costa Rica, San Jos\'e, Costa Rica.}
  
\maketitle

\begin{abstract} 
 We show that an MHD-instability driven dynamo (IDD) operating
 in a hot accretion disk is capable of generating energetically adequate
 magnetic flux deposition rates above and below a mildly advective 
 accretion disk structure.  	
 The dynamo is driven by the magnetorotational instability (MRI) of a 
 toroidal field in a shear flow and is limited by the buoyancy of 
 `horizontal' flux and by reconnection in the turbulent medium.  
 The efficiency of magnetic energy deposition is estimated to be comparable
 to the neutrino losses although an MHD collimation mechanism may deem 
 this process a more viable alternative to neutrino-burst--driven models
 of \gam-ray bursts.
\end{abstract}    


\section{Introduction}
 \label{sec:Intro}

	The combined redshift and fluence measurements of at least five
 \gam-ray burst sources plus very large photon energy detections in 
 certain bursts and tight size constraints derived from the rapid
 risetimes of burst triggers strongly suggest that the release of
 energy is highly focused by the central engine that propels
 a \gam-ray burst.  
 In spite of the very large \gam-ray energy requirements,
 the efficiency of energy deposition into 
 electromagnetic channels is likely to be very poor
 if the burst is driven by a neutrino burst in analogy to the processes
 thought to give birth to supernov\ae\, (MacFadyen \& Woosley 1998, 1999)
 or if the burst involves major energy losses to gravitational radiation
 such as might be the case in compact object merger scenarios 
 (Rasio and Shapiro 1994, Davies \etal 1994, Ruffert \& Janka 1998).
 Indeed, these measurements pose a serious energy budget problem for
 arguably all gravitational collapse powered models of \gam-ray bursts 
 if the energy release is not moderately collimated.

 	An attractive solution to 
 this problem
 starts with a (directional) Poynting-flux dominated outflow 
 (Thompson 1994,  M\'esz\'aros \& Rees 1997) under the premise 
 that such a flow may carry very little baryonic contamination 
 if deposited along a centrifugally (or gravitationally) evacuated
 funnel such as the angular momentum axis of a black hole-accretion disk 
 system.  Yet, formal motivation for an external field of
 the desired strength and topology has yet to be investigated
 in this setting.  


	We motivate a reasonable set of heuristic two dimensional
 dynamo equations for magnetic field components in the comoving frame 
 of a mildly advective disk 
 under the premise of negligible generation of meridional field
 (invoking Parker's undulate instability to promote the growth (loss) of
 vertical (horizontal) field is questionable in turbulent disks where the
 turbulence is fed by MRI's).
 A self-consistent turbulent steady state is achieved
 when the non-linear damping rate of the turbulent cascade equals the 
 inverse of the linear growth timescale 
 (Zhang, Diamond \& Vishniac 1994).
 Accretion disk models are far from being self-consistent in this respect.

 	The heuristic dynamo equations account for field generation by 
 the shear flow and by the non-axisymmetric MRI in Paczy\'nski's 
 pseudo-potential well 
 $\Phi \propto (r-r_g)^{-1}$ where $r_g = 2GM/c^2$; with  
 flux buoyancy and turbulent 
 reconnection providing for field loss terms.  
 The model predicts {\it azimuthally averaged} field components 
 and depends explicitly on the magnetic Mach number of the turbulence
 (assumed Alfvenic), 
 on local pressure ratios, and on the relativistic generalization of the 
 shear parameter (A$^{\rm Rel}_\Oor$ = Oort's A constant).

 
\section{Dynamo Equations and Scalings}
 \label{sec:DynEqu}

	In Lagrangian coordinates
\beq
 \partial_t B_\varphi 	= {B_r \over \tau_{_\cS}} 	
		- {B_\varphi \over \tau_{_\cB}}		
		- {B_\varphi \over \tau_r} 	 
~~~~~{\rm and}~~~~~
 \partial_t B_r		=  {B_\varphi \over \tau_{_\cM}} 	
		- {B_r \over \tau_{_\cB}}		
		- {B_r \over \tau_r}.	
\label{eq:EurDynEqs} \eeq

 	The shear flow is parameterized linearly by a relativistic 
 generalization of Oort's first constant (Novikov \& Thorne 1974), 
 $\tau_{_\cS}^{-1} = ~$2A$_\Oor  = \gamma^2 d_{\ln r} \Omega$,
 where $\gamma$ is the bulk Lorentz factor of the flow.
 Shear forces the radial wavenumber of perturbations to
 evolve according to $k_r(t) = k^0_r - 2{\rm A} k_\varphi t$.

 	We use exact analytical MRI scalings from 
 Foglizzo \& Tagger (1995). 
 With $\alpha_\varphi \equiv B^2_\varphi/(8\pi p)$
 and  $\AM \equiv {\rm A}/\Omega < 0$,
 the maximum growth of non-axisymmetric MRI modes  
 occurs for wavenumber 
\beq
 	k_\varphi \equiv \eta 	\, {\Omega \over \Valf_\varphi}
 	~~~~~\stackrel{\alpha_\varphi \ll 1}{\longrightarrow} 
	\{2 \AM + (1+\alpha_\varphi) \AM^2\} \, {\Omega \over \Valf_\varphi}
\eeq
 at a rate, $\tau_\cM^{-1}$, that obeys 
\beq
	| \AM |^2 \tau^2_{_\cM} 
 	\stackrel{\alpha_\varphi \ll 1}{\longrightarrow} 
	1 + \alpha_\varphi ( 2 + \AM)
\eeq
 as long as $k_\varphi \gtaprx 1/r$. On the other hand,
 in a very strongly sheared flow
\beq 	-\AM > 1 + (2\alpha_\varphi+1)^{-1},
\label{eq:RadInt} \eeq 
 the slow branch of MHD propagation
 (to which the toroidal MRI belongs) is destabilized into
 a radial interchange mode (T.~Foglizzo, Priv. Comm.) 
 in accordance with the Rayleigh criterium.  
\pagebreak
\pagebreak
 
	A plausible mechanism that promotes baryon unloading from field 
 lines is turbulent pumping (Vishniac 1995a) by the MRI which must
 favorably compete with turbulent diffusion of matter back onto flux ropes.  
 Under this assumption, the stretch, twist and fold of field lines
 by (enthalpy-weighted) sub-Alfvenic turbulence augments the field 
 energy density and releases matter from field lines that 
 otherwise would be ``frozen-in".  
 For the marginal case of Alfvenic turbulence,
 nearly empty B$_\perp$ flux ropes (i.e. flux residing on surfaces
 perpendicular to the local meridian) in a gas pressure
 dominated disk acquire a drag limited buoyant velocity 
 $v_b \propto (\Valf_{\perp\theta})^2/c_s$. 
 Moreover, assuming efficient diffusion of radiation and
 $e^\pm$ pairs into the flux tubes, in this picture 
 the buoyancy loss rate, $\tau_\cB^{-1} = v_b / \cH_\theta$, is
 enhanced by a factor $\ltaprx ~ p/\Pgas \equiv \xi$ (Vishniac 1995b). 

 	Reconnection of the field at sub-MRI optimal lengthscales
 (where the field lines are only weakly stochastic) occurs 
 at the \Alf speed for Alfvenic turbulence (Lazarian \& 
 Vishniac 1999).  The rates may be written as 
 inverse \Alf transit times calculated from the
 component of the field that undergoes reconnection, 
 $\tau_{\rm rec} \approx {l^{\perp i}_{\rm rec} / \Valf_i}
 = \sqrt{2 / \Gamma} \, ({l^{\perp i}_{\rm rec} / \cH_\Theta}) \,
		({c_s / \Valf_i}) \; \Omega^{-1}$,
 where $\Gamma$ is the adiabatic index of the fluid.

 	The perpendicular lengthscales, $l^{\perp i}_{\rm rec}$,
 associated with the mean distance for field reversal are derived from 
 the fundamental linear lengthscales for coherent field pumping.
 These are supplied by half of the toroidal MRI's (wave)length
 scale.  Azimuthal, radial field reversal, $Y_r$, directly involves the
 optimal wavenumber of the non-axisymmetric MRI.  In addition, following
 Tout \& Pringle (1992), 
 an estimate of the radial, azimuthal field reversal lengthscale, 
 $X_\varphi$ follows by noting that the time evolution of wavenumbers 
 implied by shear during one MRI timescale
 couples the azimuthal lengthscale to the radial lengthscale, 
 i.e. $l_y^\perp = (\tau_{_\cM}/\tau_{_\cS}) \times  l_x^\perp$.
 Thus $	Y_r = \pi /{k_{_\cM}}~{\rm and}~
	X_\varphi = ( {\tau_{_\cS} / \tau_{_\cM}} ) \times 
	(\pi / {k_{_\cM}})$.

\section{Equilibrium Solutions} 
 \label{sec:Equil_Sol} 

	Scaling wavenumbers to the inverse pressure scale height 
 $k \rightarrow \hat{k} / \cH_\Theta$, 
 and {\it redefining} 
 $\AM \equiv |{\rm A}|/\Omega > 0$;
 a set of normalized dynamo equations follows 
 by representing fields in (\Alf) velocity units and normalized to the 
 soundspeed ($B' = B \times \sqrt{4\pi\varrho} \, c_s$), and the time
 normalized to the inverse of the Keplerian frequency $t' = \Omega t$.

	In a steady state, these equations must satisfy
\beq
 B'_r 	= { B'_\varphi \over {2 \AM} } 
	\,\left\{ {1 \over \xi} \,  {B'_\perp}^{\!\!2} 
	+ {\sqrt{2}\over\pi} \, {\tau'_{_\cM} \over \tau'_{_\cS}} {\eta} 
 \right\} , 
~~{\rm and}~~
 B'_\varphi 	= { B'_r \over {\AM} } 
	\,\left\{ {1 \over \xi}  {B'_\perp}^{\!\!2} 
	~+~ {\sqrt{2}\over\pi} \, {B'_r \over B'_\varphi} {\eta} 
 \right\} 
\eeq
 which we solve numerically.

\section{The energy deposition rate}
 \label{sec:results}

 The accretion disk setting is envisioned to 
 follow the standard hyper-accreting black hole model of 
 Popham, Woosley \& Fryer (1999, hereafter PWF) where
 $M_\bh = 3 \msun; ~\alpha_\SS = 0.1, ~{\rm and}~ 
 \dot{M} = 0.1 \msun \sec^{-1}$.
 By adopting their published pressure ratios 
 and assuming Keplerian rotation for 
 $r \in [2.25, ~20]r_g$, we ``piggyback" 
 the hydromagnetic energy conversion process on this model. 

	We find that 
 the magnetic output rate from buoyancy,
 $\dot{E} \simeq 1.77_{+51}$ erg sec$^{-1}$, 
 is comparable to the neutrino luminosity 
 $L_\nu \simeq 3.3_{+51}$ erg sec$^{-1}$ (PWF).
 The `half-luminosity' radius is located at $r_{\cal L} \approx 5.75 \, r_g$
 and the IDD becomes operational 
 (against the radial interchange instability, c.f. Eq [\ref{eq:RadInt}]
 and Araya-Gochez 1999) at $r_{\rm min} \simeq 2.55 \, r_g$.  
 Curiously, the derived value of the magnetic viscosity 
 $\alpha_\SS = B'_r B'_\varphi$ hovers on 0.1 at the innermost radii
 and decreases to about .087 at $r_{\cal L}$ in good agreement with the 
 adopted value.   
 	Lastly, we note that the MRI pumps the field preferably at the 
 lowest wavenumbers for a near-equipartition field in a strongly sheared 
 flow.  Thus, most of this energy could in principle  
 go into a Poynting jet with interesting consequences for \gam-ray bursts.     


\end{document}